\documentclass[12pt,preprint]{aastex}

\usepackage{natbib}
\usepackage[usenames]{color}

\newcommand{\lsi}{LS~I~$+$61$^{\circ}$303}

\shorttitle{Long-term $\gamma$-ray variability of \lsi}
\shortauthors{The {\it Fermi}-LAT collaboration}

\begin{document}

%%%%%%%%%%%%%%%%%%%%%%%%%%%%%%%Title%%%%%%%%%%%%%%%%%%%%%%%%%%%%%%%%%%%%%
\title{Associating long-term $\gamma$-ray variability \\ with the superorbital period of \lsi }

\author{
M.~Ackermann$^{1}$,
M.~Ajello$^{2}$,
J.~Ballet$^{3}$,
G.~Barbiellini$^{4,5}$,
D.~Bastieri$^{6,7}$,
R.~Bellazzini$^{8}$,
E.~Bonamente$^{9,10}$,
T.~J.~Brandt$^{11}$,
J.~Bregeon$^{8}$,
M.~Brigida$^{12,13}$
% ...
}

\email { {  \dag} A. Caliandro (andrea.caliandro@ieec.uab.es), D. Hadasch (hadasch@ieec.uab.es), D. F. Torres (dtorres@ieec.uab.es)}

\altaffiltext{1} {Deutsches Elektronen Synchrotron DESY, D-15738 Zeuthen, Germany}
\altaffiltext{2} {Space Sciences Laboratory, 7 Gauss Way, University of California, Berkeley, CA 94720-7450, USA}
\altaffiltext{3} {Laboratoire AIM, CEA-IRFU/CNRS/Universit\'e Paris Diderot, Service d'Astrophysique, CEA Saclay, 91191 Gif sur Yvette, France}
\altaffiltext{4} {Istituto Nazionale di Fisica Nucleare, Sezione di Trieste, I-34127 Trieste, Italy}
\altaffiltext{5} {Dipartimento di Fisica, Universit\`a di Trieste, I-34127 Trieste, Italy}
\altaffiltext{6} {Istituto Nazionale di Fisica Nucleare, Sezione di Padova, I-35131 Padova, Italy}
\altaffiltext{7} {Dipartimento di Fisica e Astronomia ``G. Galilei'', Universit\`a di Padova, I-35131 Padova, Italy}
\altaffiltext{8} {Istituto Nazionale di Fisica Nucleare, Sezione di Pisa, I-56127 Pisa, Italy}
\altaffiltext{9} {Istituto Nazionale di Fisica Nucleare, Sezione di Perugia, I-06123 Perugia, Italy}
\altaffiltext{10} {Dipartimento di Fisica, Universit\`a degli Studi di Perugia, I-06123 Perugia, Italy}
\altaffiltext{11} {NASA Goddard Space Flight Center, Greenbelt, MD 20771, USA}
\altaffiltext{12} {Dipartimento di Fisica ``M. Merlin" dell'Universit\`a e del Politecnico di Bari, I-70126 Bari, Italy}
\altaffiltext{13} {Istituto Nazionale di Fisica Nucleare, Sezione di Bari, 70126 Bari, Italy}
% ...
%\altaffiltext{*] current address

%%%%%%%%%%%%%%%%%%%%%%%%%%%%%%% Abstract %%%%%%%%%%%%%%%%%%%%%%%%%%%%%%%%%
\begin{abstract}

Gamma-ray binaries are stellar systems for which the spectral energy distribution 
(discounting the thermal stellar emission) 
peaks at high energies. Detected from radio to TeV gamma rays, the $\gamma$-ray binary \lsi\ is highly variable across all frequencies. One aspect of this system's variability is the modulation of 
its emission with the timescale set by the $\sim 26.4960$-day orbital period. 
Here we show that, during the time of our observations, the $\gamma$-ray emission of \lsi\ also presents
a sinusoidal variability consistent with the previously-known superorbital period of 1667 days.
This  modulation  is more prominently seen  at orbital phases around apastron, whereas it does not introduce a visible change close to periastron. It is also found in the appearance and disappearance of variability at the orbital period in the power spectrum of the data.
This behavior could be explained by a quasi-cyclical evolution of the equatorial outflow of the Be companion star, whose features influence the conditions for generating gamma rays. 
These findings open the possibility to use $\gamma$-ray observations to study the outflows of massive stars in eccentric binary systems.

\end{abstract}

%%%%%%%%%%%%%%%%%%%%%%%%%%%%%%%%% Section 1 %%%%%%%%%%%%%%%%%%%%%%%%%%%%%%%
\section{Introduction}

\lsi\ is one of the few X-ray binaries that have been detected from radio to TeV gamma rays (see Albert et al. 2006 and references therein).
It is perhaps the most intriguing one due to the high variability and richness of its phenomenology at all frequencies. 
\lsi\  consists of a Be star of approximately 10 solar masses, and a compact object. %(see e.g., \cite{Casares2005}). 
Be stars are rapidly rotating B-type stars showing hydrogen Balmer lines in emission in the stellar spectrum, and which lose mass to an equatorial circumstellar disc. 
The nature of the compact object in \lsi\ has been much debated over the past few years:
Pulsar wind interaction (see e.g., \citealp{Maraschi1981,Dubus2006,Zamanov2001,Torres2012}) and microquasar jets (see \citealp{Bosch-Ramon2009} for a review) have been proposed as the origin of the non-thermal emission. 
The recent detection of two short ($<0.1 $ s), highly-luminous ($>10^{37}$ erg s$^{-1}$), thermal flares \cite{Papitto2012}
have given support to the hypothesis
that the compact object in \lsi\ is a neutron star, for only highly-magnetized neutron stars have been found to behave in this way.

The flux of \lsi\ is seen to be modulated by the orbital period of 26.4960 days  \cite{Gregory2002}  at most wavelengths, including at high energies 
\cite{Torres2010,Zhang2010, Abdo2009,Albert2008}. 
Orbital modulation of the GeV flux can be understood as a consequence of changing conditions for generation and absorption of gamma rays, which are mostly determined by the orbital geometry; e.g., the viewing angle to the observer and the position of the compact object with respect to the stellar companion. Unless other physical conditions change, we do not expect long-term variability of the emission level at a fixed orbital configuration. 
In order to investigate \lsi's variability, 
we analyzed {\it Fermi}-Large Area Telescope (LAT) data
from the beginning of scientific operations on {2008 August 4} until {2013 March 24}.
We report on the results in this {\it Letter}.

\section{Data Analysis}

%The Large Area Telescope (LAT) is the primary instrument onboard the {\it Fermi} Gamma-ray Space Telescope satellite; it is a pair-conversion telescope sensitive to photons with energies from $\sim$20\,MeV to more than 300\,GeV.
%The observatory operates principally in survey mode; the telescope is rocked north and south on alternate orbits to provide more uniform coverage so that every part of the sky is observed for $\sim$30 minutes every 3 hours.
%The analysis presented here makes use of all data taken from the beginning of scientific operations on {2008 August 4} until {2013 March 24}.
%The data were analyzed using the LAT {Science Tools} package (v9r30),
%which is available from the {\emph{Fermi}} Science
%Support Center.

We used the LAT {Science Tools} package (v9r30),
which is available from the {\emph{Fermi}} Science
Support Center,  as is the LAT data, together with the {\emph{P7v6}} version of the
instrument response functions. 
%
%The analysis used the {\emph{P7v6}} version of the instrument response functions.
%which take into account accidental coincidence effects in the detectors.
%
Only events passing the Pass 7 ``Source'' class cuts are used in the analysis.
All gamma rays with energies $>100$ MeV
within a circular region of interest (ROI) of 10$^{\circ}$ radius centered on \lsi\
were extracted.
%The good time intervals are defined such that source is always inside the LAT field of view, namely within a cone angle of 66$^{\circ}$. 
To reduce the contamination from the Earth's upper atmosphere time intervals when the
Earth limb was in the field of view were excluded, specifically when the
rocking angle of the LAT was greater than $52^{\circ}$ or when parts
of the ROI were observed at zenith angles
$>100^{\circ}$.
The $\gamma$-ray flux of \lsi\    plotted in the light curves of this work are calculated by performing the binned 
%Fig 2
or the unbinned 
%(Fig. 1,3) 
maximum likelihood method, %\cite{Mattox1996}, 
depending on the statistics,
by means of the Science Tool {\tt gtlike}.
The spectral-spatial model constructed to perform the likelihood analysis includes
all the sources of the second {\it Fermi}-LAT point-source catalog \cite{2FGL} (hereafter 2FGL) within 15$^\circ$ of \lsi.
The spectral parameters were fixed to the catalog values, except for the sources within 3$^\circ$
of \lsi. For these latter sources, the flux normalization was left free.
\lsi\ was modeled with an exponentially cut off power-law spectral shape.
All its spectral parameters were allowed to vary (see \citealp{Hadasch2012} for further details).
The models adopted for the Galactic diffuse emission (gal\_2yearp7v6\_v0.fits) and isotropic backgrounds (iso\_p7v6source.txt) were those  recommended by the LAT team.\footnote{A description of these models is available from the {\it Fermi} Science Support Center: {\url http://fermi.gsfc.nasa.gov/ssc/data/access/lat/BackgroundModels.html}.}

Systematic errors mainly originate in the uncertainties in the effective area of the LAT, as well as in the Galactic diffuse emission model.
The current estimate of the uncertainties of the effective area is 10\% at 100 MeV, decreasing to 5\% at 560 MeV and increasing to 10\% at 10 GeV and above. We assume linear extrapolations, in log space, between the quoted energies.
%For details on the systematic uncertainties of {\emph Fermi}-LAT see \cite{2FGL,sys}.
The systematic effect is estimated by repeating the likelihood analysis using modified instrument response functions that bracket the ``{\tt P7SOURCE\_V6}'' effective areas.\footnote{The released Pass 7 Instrument Response Functions are documented here: {\url {http://www.slac.stanford.edu/exp/glast/groups/canda/lat\_Performance.htm}.}}
Specifically, they are a set of Instrument Response Functions in which the effective area has been modified considering its uncertainty as a function of energy in order to maximally affect a specific spectral parameter.
In order to conservatively take into account the effect due to the uncertainties of
the Galactic diffuse emission model, the likelihood fits are repeated changing the
normalization of the Galactic diffuse model artificially by $\pm 6\%$. We have found flux systematic errors (for energies above 100 MeV) 
on the order of 9\%, similar to the ones reported
in \citealp{Hadasch2012}.

\section{Results}

\begin{figure}[!t]
\centering
\includegraphics[width=0.9\textwidth]{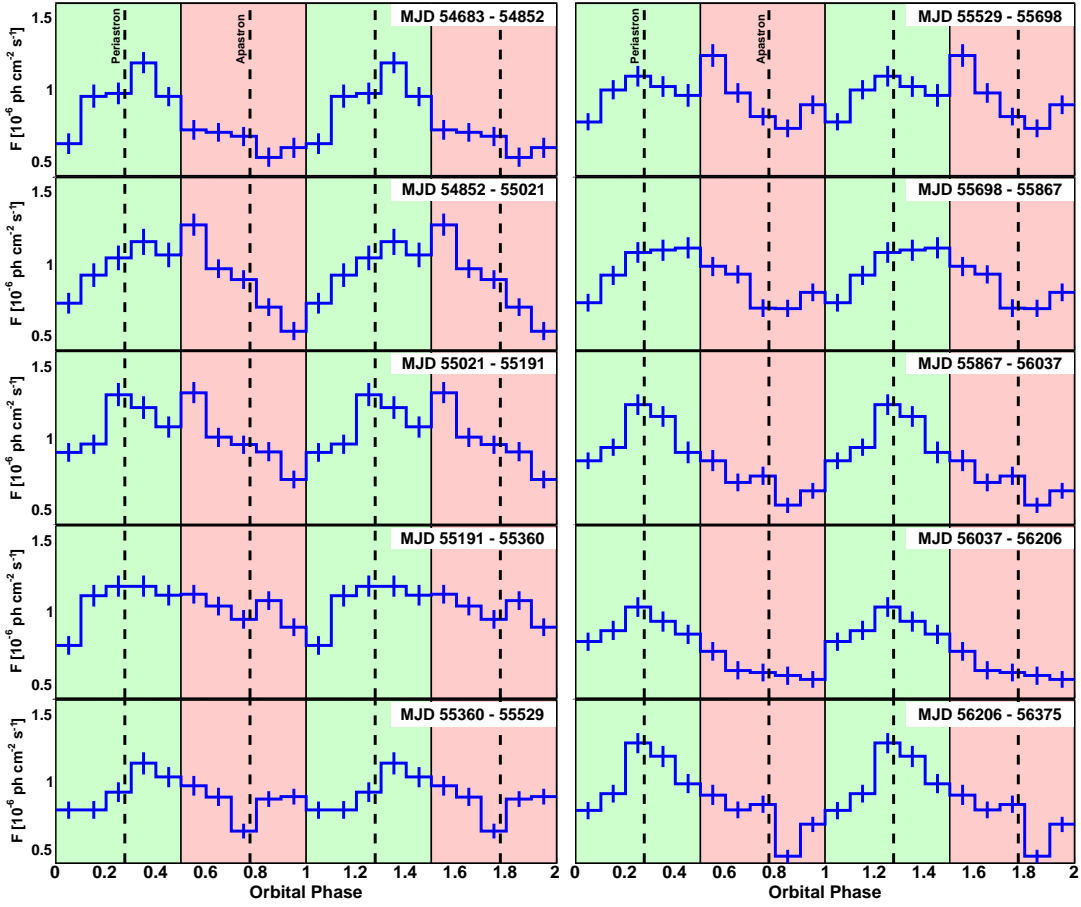} 
\caption{Gamma-ray flux from \lsi\ folded on the orbital period. The data are repeated over two cycles for clarity.
Photons with energies above 100 MeV, as measured by {\it Fermi}-LAT are considered.
The measurements cover the period from {2008 August 4} to {2013 March 24}, from the top left panel to the bottom right.
Each panel  spans an equal interval of {169.2} days. The position of periastron and apastron are marked with
dashed vertical lines 
 (the ephemeris
of \citealp{Aragona2009} is used). The two background colors correspond to the periastron (orbital phases 0.0--0.5) 
and apastron (orbital phases 0.5-1.0) regions of the orbit. } 
\label{orb-var}
\end{figure}

Fig.~\ref{orb-var} shows the  orbitally-folded light curve of \lsi\
from {2008 August 4} to {2013 March 24}. 
It shows a trend for  the maximum of the $\gamma$-ray emission 
to appear 
near periastron (phases around 0.3), as in \citealp{Hadasch2012}, and
significant $\gamma$-ray flux variability at fixed orbital phases. 
%Taking into account the errors in the measurements,  the variability is at a level of 
%4 to 8$\sigma$ at phases 0.5 to 1.0, and it is at least at the level of $\sim 3\sigma$ at others orbital bins. \\

\begin{figure}[t!]
\centering
\includegraphics[width=0.7\textwidth]{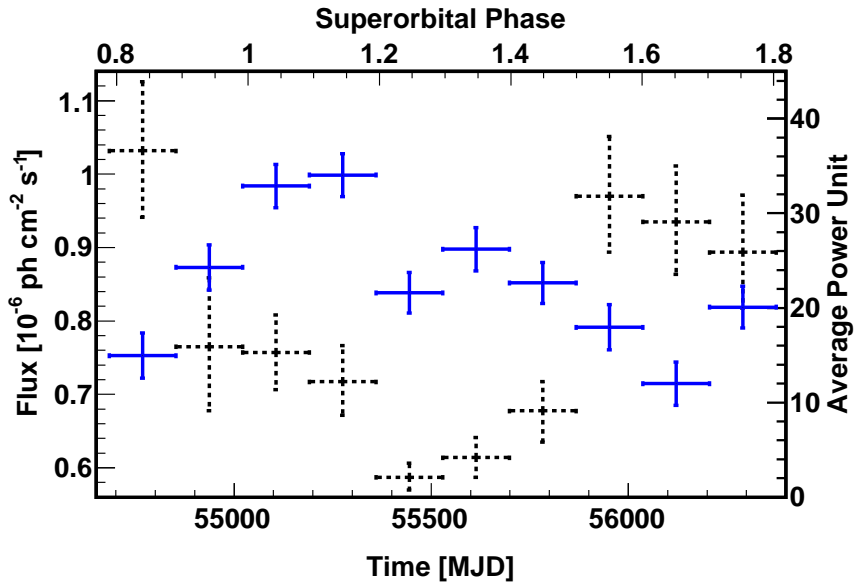} 
\caption{Long-term evolution of the average $\gamma$-ray flux  (above 100 MeV) from \lsi\ (blue points, left y-axis scale). The 
superorbital phase is shown in the top axis. The right y-axis scale and the black dashed points
show the long-term evolution of the power at the orbital period found in the Lomb-Scargle periodogram.
%A significance of 3$\sigma$ (5$\sigma$) corresponds to about 6 (15) Average Power Units. 
 } 
\label{flux-so}
\end{figure}

We explore the possibility that the observed long term $\gamma$-ray variability could  be related to the superorbital period of 1667$\pm$8 days as reported in radio and optical frequencies
\cite{Gregory2002}. A variability signature with this period was also found along several years of X-ray observations \cite{Li2012,Chernyakova2012}. 
Fig.  \ref{flux-so} shows the long-term evolution of the average $\gamma$-ray flux; we use the superorbital period of Gregory (2002) to translate time to superorbital phase.
The probability that this evolution is a random result out of a uniform distribution is {$<1.1\times 10^{-12}$} ($\chi^2$, $ndf$ = 75.8, 9).

\begin{figure*}
\centering
\includegraphics[width=0.9\textwidth]{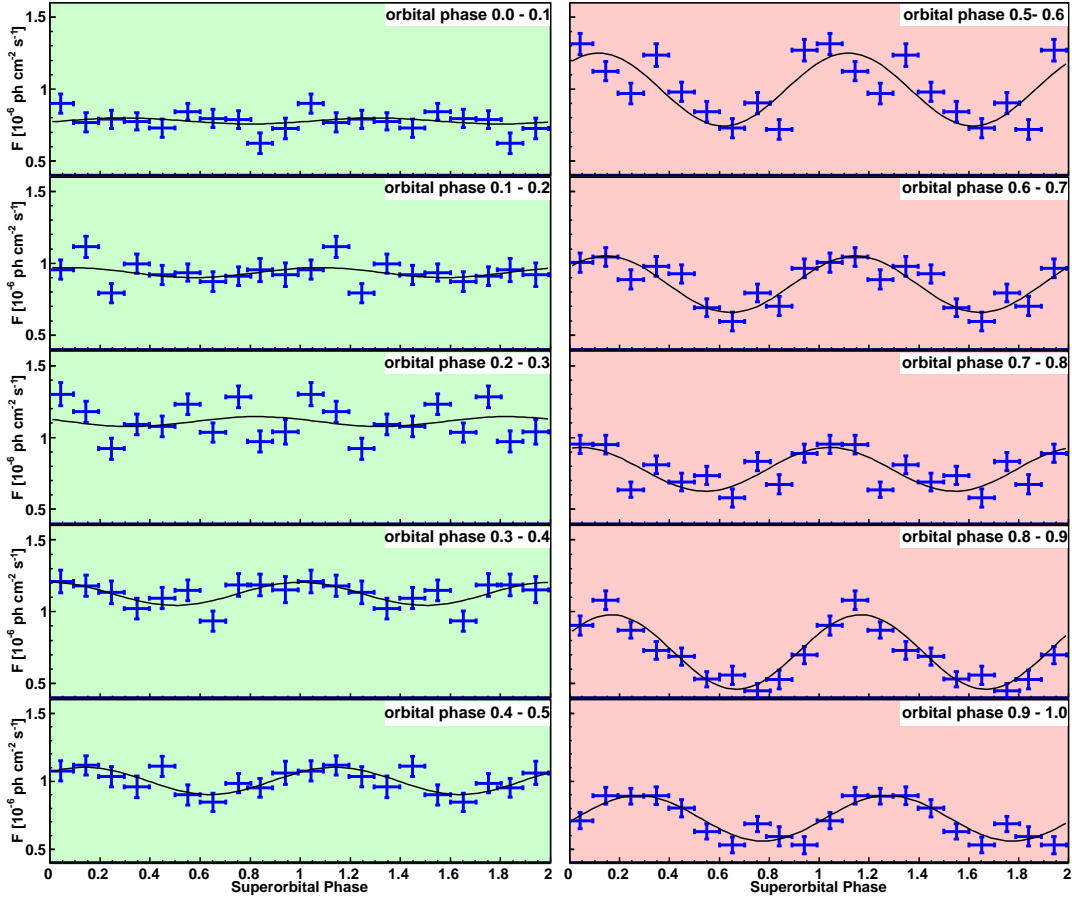} 
\caption{The evolution of the $\gamma$-ray flux (above 100 MeV) from \lsi\ at fixed orbital phases as a function of the superorbital phase. 
The data points are repeated over two superorbital periods for the sake of clarity. The left panels
represent the region of the orbit near periastron (located at phase $\sim 0.3$, see Fig.  \ref{orb-var}) where
the data are compatible with no superorbital variability beyond 3$\sigma$.). The right panels, instead,
are regions close to apastron. The black line in each of the panels is a sinusoidal function fit to the
data points, with a fixed period of 1667 days.} 
\label{superorb-var}
\end{figure*}

To check for a possible long-term modulation of the $\gamma$-ray flux at any orbital configuration, 
we have separated 
the data in orbital bins, and plotted
the fluxes against the superorbital phase, as shown in  Fig.  \ref{superorb-var}.
The black line in each of the panels of Fig.  \ref{superorb-var} represents a sinusoidal function fit
to the data points. The period of this function has been kept (in all panels) at the value of the superorbital
period found in radio (1667 days). 
Thus, the function we use to fit the data has three parameters: average flux level, amplitude, and phase. 
We have also fitted a constant line for comparison. 

Table \ref{fit-results} shows the quality of the fitting results corresponding to Fig.  \ref{superorb-var}. 
It has the following columns: the system's orbital phase,
the corresponding $\chi^2$ and {\it dof}
as well as 
the probability that the data are described by either a constant or a sinusoidally varying flux,
and finally the probability that the improvement
found when fitting a sinusoid instead of a constant is produced by chance. To obtain the latter we consider the likelihood ratio test \cite{Mattox1996}. The test is performed by computing
the ratio $2\times\Delta log(Likelihood)$ for the two hypotheses (constant and sinusoidal) and assuming that for a chance coincidence
the ratios are $\chi^{2}$-distributed according to the difference in the degrees of freedom between the two hypotheses.
Thus, if the hypothesis of a constant is true, the likelihood ratio $R = -2 \ln(L(constant)/L(sine))$ is approximately $\chi^2$-distributed with 2 degrees of freedom.
%where 2 is the difference between the number of the free parameters of two hypotheses. 
The probability that one hypothesis is preferred over the other is defined as 
$P = \int_0^{R_{meas}} p(\chi^2) d\chi^2$
where $p(\chi^2)$ is the $\chi^2$ probability density function and $R_{meas}$ the
measured value of $R$. The constant hypothesis will be rejected (and the sinusoidal will be accepted) if $P$ is greater than the confidence level, which is set to 95\%. In Table \ref{fit-results}, the last column states the probability that the fit improvement (of a sine over a constant) is happening by chance (thus, $1-P$).

\begin{table*}
\scriptsize
\caption{Quality of the fitting results corresponding to Fig.  \ref{superorb-var} (top panel) and sinusoidal fitting parameters for the flux near apastron (bottom panel).
%For details on the columns shown, see text. 
}
\begin{center}
\begin{tabular}{l r r r r r}
\hline
\hline
Orbital & $\chi^2$, ndf & Constant Fit &  $\chi^2$, ndf & Sine Fit & Prob. improvement \\
Phase &  (constant) & Probability &  (sine) &Probability& by chance \\
\hline
0.0--0.1 & 10, 9 &  $3.2 \times 10^{-1}$ & 10, 7 & $1.9 \times 10^{-1}$ & 1.0  \\
0.1--0.2 & 13, 9 & $1.8 \times 10^{-1}$  & 12, 7 & $1.1 \times 10^{-1}$ &1.0 \\
0.2--0.3 &  27, 9 & $1.4 \times 10^{-3}$ &  26, 7 & $5.0 \times 10^{-4}$  & 0.7 \\
0.3--0.4 &  13, 9 & $1.6 \times 10^{-1}$ &  8, 7 & $3.6 \times 10^{-1}$ & $7.0 \times 10^{-2}$\\
0.4--0.5 &  15, 9 & $9.9 \times 10^{-2}$ &  6, 7 & $5.4 \times 10^{-1}$ & $1.2 \times 10^{-2}$\\
0.5--0.6 &  84, 9 & $2.8 \times 10^{-14}$ & 23, 7 & $2.0 \times 10^{-3}$ & $<1.0 \times 10^{-7}$\\
0.6--0.7 &  50, 9 & $8.1 \times 10^{-8}$ &  10, 7 & $2.2 \times 10^{-1}$ & $<1.0 \times 10^{-7}$\\
0.7--0.8 &  41, 9 & $6.1 \times 10^{-6}$ &  18, 7 & $1.0 \times 10^{-2}$ & $1.4 \times 10^{-5}$\\
0.8--0.9 & 100, 9 & $2.4 \times 10^{-17}$ &  8, 7 & $3.0 \times 10^{-1}$ & $<1.0 \times 10^{-7}$\\
0.9--1.0 &  50, 9 & $9.1 \times 10^{-8}$ &  10, 7 & $2.2 \times 10^{-1}$ & $<1.0 \times 10^{-7}$\\
\hline
\hline
Orbital & $F_0$ & $A$ & $\phi$ \\ 
Phase & [$10^{-6}$ ph cm$^{-2}$ s$^{-1}$] & [$10^{-6}$ ph cm$^{-2}$ s$^{-1}$]  \\
\hline
0.5--0.6 & 1.00$\pm$0.03  & 0.25$\pm$0.03  &  0.87$\pm$0.03\\
0.6--0.7 & 0.85$\pm$0.02  & 0.20$\pm$0.03  &  0.90$\pm$0.02\\
0.7--0.8 & 0.78$\pm$0.02  & 0.15$\pm$0.03  &  0.79$\pm$0.03\\
0.8--0.9 & 0.72$\pm$0.03  & 0.26$\pm$0.03  &  0.92$\pm$0.03\\
0.9--1.0 & 0.73$\pm$0.02  &  0.17$\pm$0.03  &  0.02$\pm$0.04\\
\hline
\hline
\end{tabular}
\end{center}
\label{fit-results}
\end{table*}%

Table \ref{fit-results} also shows the sinusoidal fit parameters
corresponding to the right-hand panels of Fig. \ref{superorb-var}. The functional form of the fit
is $F_0 + A \times \sin(  (t-T_0) / T -\phi)\times2\pi)$. Here, $T_0$ and $T$ are the zero time ($T_0$ = MJD 43366.275) and the period (always kept fixed at 1667 days in all panels) of the superorbit, respectively (both as in \citealp{Gregory2002}), $t$ is the time,
$F_0$ is the average flux level, $A$ is the amplitude, and $\phi$ represents the phase shift in the superorbit.
The choice of a sinusoidal function for fitting the data is not based on any {\it a priori}
physical expectation; the superorbital variability could be periodic but have a different shape. However, any periodic function
could be described by a series of sines.
Thus, fitting with just one sinusoidal function 
as  done above is motivated by the relatively low number of data points.

No strong variability is found at orbital phases 0.0--0.5, while it is clearly present in the range 0.5--1.0.
Concurrently, data at the orbital phases 0.0 to 0.5 are not significantly 
better-represented by a sine than by a constant. However, this is not the case for the data at the orbital phases 0.5 to 1.0. 
The probability that the sinusoidal fit improvement occurs by chance is less than 
$1.0\times 10^{-7}$ at orbital phases 0.5--0.6, 0.6--0.7, 0.8--0.9, and 0.9--1.0; and $1.4\times 10^{-5}$ at orbital phases 0.7-0.8.
Whereas the sinusoidal variation is always a better fit in this part of the orbit, the amplitude of the fit is maximal 
in orbital phases before and after the apastron.

\begin{figure*}
\centering
\includegraphics[width=0.45\textwidth]{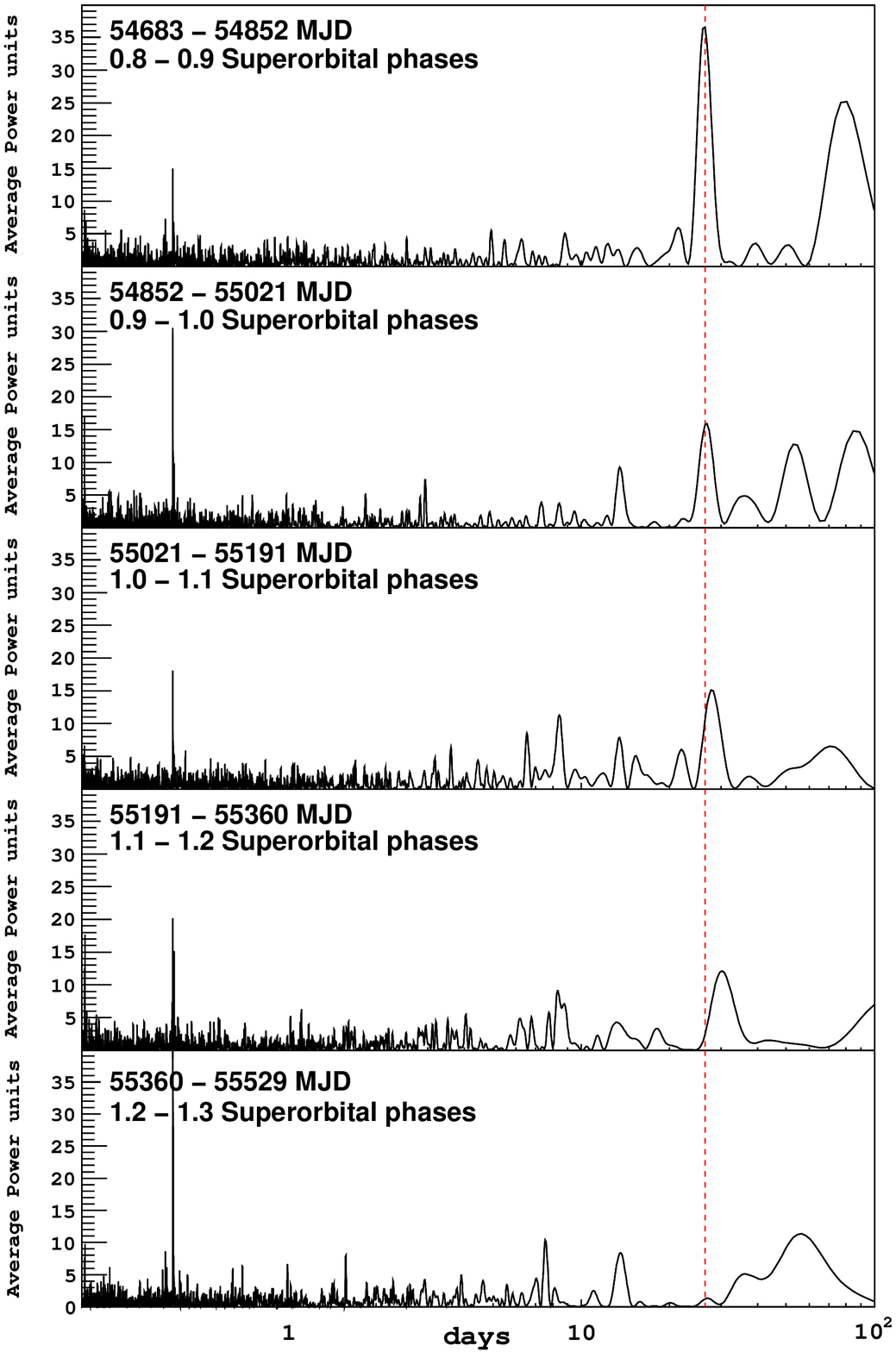}
\includegraphics[width=0.45\textwidth]{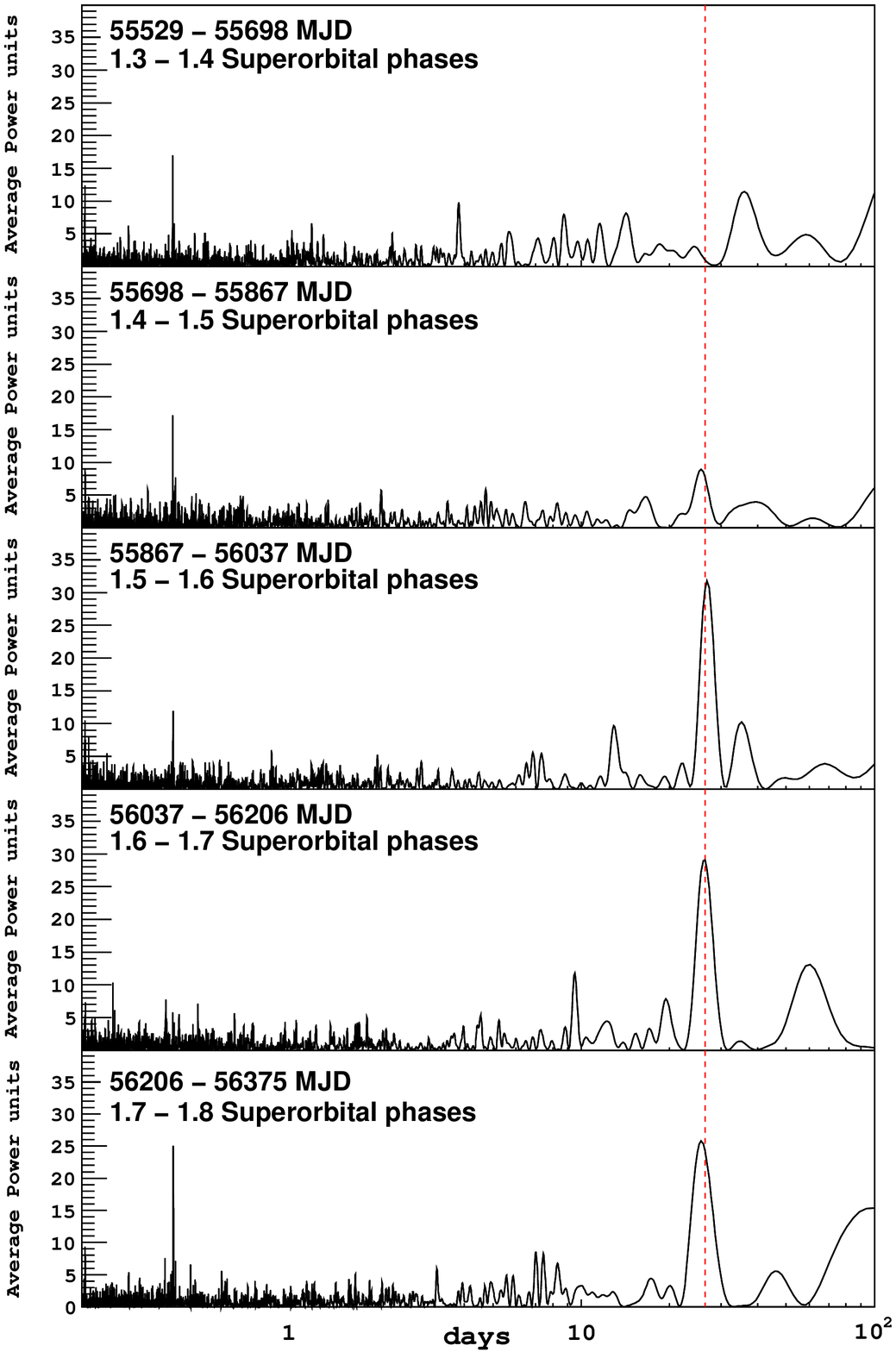} 
\caption{Periodogram of the $\gamma$-ray data for different time intervals. The dashed line marks the orbital period
of \lsi.
 } 
\label{powspec}
\end{figure*}

In order to test for the appearance/disappearance of the orbital signature in gamma rays,
we subdivided the data into the same time intervals of Fig.  \ref{orb-var} 
%(intervals of 169.2 days each), 
and applied the Lomb-Scargle periodogram technique \cite{Lomb1976,Scargle1982} to each of them. 
To calculate the power spectrum the event selection was restricted to a ROI of 3$^{\circ}$ radius centered on \lsi.
%, and the good time intervals were recalculated. 
The selected events were used to create a light curve of weighted counts over exposure with equally spaced time bins of 2.4 hours width. The weight associated to each event corresponds to the probability that the $\gamma$-ray was emitted by \lsi, rather than by nearby sources or has a diffuse origin
%\cite{Kerr2011}.
The weights are calculated using the {\it Science
  Tool}  {\tt gtsrcprob}, adopting the best spectral-spatial models obtained by the binned likelihood fits described in the previous section.
%The exposure calculated for each time bin is the product of the time the source is in the field of view times the instrument effective area.
Before calculating the power spectrum, we also applied to the light curve the exposure weighting described in \citealp{Corbet2007}.
Fig.  \ref{powspec} shows the power spectra calculated in each of the time intervals.
The vertical line marks the orbital period (as in \citealp{Gregory2002}).
The y-axis in the periodograms is given in average power units, which converts the original
spectrum in units of (ph cm$^{-2}$ s$^{-1}$)$^2$ by normalizing it with the average of the power over all the frequencies $<P>$. In this way, the units are directly linked to the significance of the peak, which for a peak of power $\bar P$ is computed as ${\rm Prob}(P>\bar P) = \exp {(- \bar P / <P>)}$ \cite{Scargle1982}. These average power values are plotted in Fig. \ref{flux-so}.
A significant peak is detected at the orbital period, but not in all time intervals.
Note that in some of the panels of Fig.  \ref{powspec} there appears to be a shift of the 26.5-day peak, even though
it is within the fundamental frequency ($1/T_{obs}$) of the orbital period.  
A claim that the period shift of these peaks is significant 
would then imply a severe oversampling of the Fourier resolution, which for the duration of this dataset is 3.84 days.
%$DP_{Fourier}=P^2/(T_{obs})=(26.5 {\rm days})^2/(0.5 \times 365.25$ days)=3.84 days.
The shifted peaks are not significant
either in the single-trial (looking for an specific frequency) or in the all-trials probability analysis of these power spectra.
%
%Our previous analysis of the first 2.5 years of LAT data \cite{Hadasch2012} has shown
%that the power spectrum peak due to the orbital flux modulation disappeared as time advanced. 
Thus, we have now found that along the time covered by our observations,
the power spectrum peak 
at the orbital period
is significant only at superorbital phases 
$\sim 0.5-1.0$. At other superorbital phases, the peak  is
absent or has a significance less than 3$\sigma$. 

\section{Discussion}

Over the last two decades, systematic
monitoring of many Be X-ray systems allowed the discovery of many cases of 
superorbital cycles (see, e.g., \citealp{Alcock2001}, \citealp{Rajoelimanana2011}). Thus, in order to connect the discovered $\gamma$-ray observational pattern to conditions that vary over the superorbit, 
a quasi-cyclical expansion and shrinking of the circumstellar disc of a Be star may offer an alternative  (e.g., \citealp{Negueruela2001}).
The sizes of the stellar discs of Be stars are hypothesized to correlate with the equivalent width (EW) of the H$\alpha$ 
emission line (e.g., \citealp{Grundstrom2006}). 
In the longest-running campaign observing \lsi\,
the maximum of the H$\alpha$ EW has been found in a broad region around superorbital phase 
{  $0.2$} (see \citealp{Zamanov1999,Zamanov2000} and references therein).
%Paredes1994}).
%
Thus, the X-ray \cite{Li2012} as well as the $\gamma$-ray
emission are enhanced at superorbital phases where  maximal values of the H$\alpha$ EW have been measured. %\cite{note3}. 
Concurrently, the power spectrum peak at the orbital period is less significant. 
This suggests that the disc may play a role in modulating both the gamma and the X-ray signals.

From the results in Fig.  \ref{superorb-var}, one may conclude that in the periastron region, when 
the  emission from the system  is subject to essentially no superorbital variability, the conditions for the 
generation of gamma rays in the GeV range
must not significantly change. We can thus assume that the compact
object could be inside or severely affected by the Be disc matter when it is closer to the companion star (i.e., at orbital phases 0.0 to 0.5),  for all superorbital phases. 
If this is the case, 
even when the EW of the H$\alpha$  line (and thus the radius within which the disc influences)
changes by a factor of a few along the superorbital period \footnote{The mass-loss rate variations from the Be star in \lsi\ were estimated  as the ratio between maximal and minimal values of its radio emission (a factor of $\sim$5 was determined by \cite{Gregory1989,GregoryNeish2002}) or its H$\alpha$ measurements, which span factors of $\sim$1.5 to 5 \cite{Zamanov1999,Grundstrom2007,Zamanov2007,McSwain2010}.}, this does not necessarily 
imply a  significant change in the $\gamma$-ray modulation above the sensitivity of {\it Fermi}-LAT when the compact object is near periastron.
However, in a two-component model typically assumed 
for Be stellar winds (an equatorial wind generating the disc, and a polar outflow) 
the conditions in the 
apastron region (e.g., the pressure exerted by the wind, or the mass gravitationally captured by the compact object) could change by more than 3 orders of magnitude if one or the other component dominates (see, e.g., \citealp{GregoryNeish2002} and references therein).
In such a case, it is reasonable to suppose that the GeV emission would be affected at an observable level.

We notice from Fig.  \ref{superorb-var}
that between the orbital phase ranges 0.9--1.0 and 0.0--0.1 there is a significant change of the long-term behavior
of the $\gamma$-ray emission. Closer to periastron the flux evolution flattens.
We can then estimate the radius at which the matter in the disc of the Be star
produces a stable influence with time by computing the system separation at orbital phase $\sim0.1$.
Using the ephemeris given by Aragona et al. (2009), we obtain a separation of $\sim 9 R_s$, where $R_s$
is the stellar radius of the Be star.
On the other hand, from the fact that the maximal amplitude of the superorbital variability
is before and after the apastron of the system, the system separation at orbital phases 0.7 and 0.9 ($\sim 13 R_s$) could
also have a physical meaning. It is a qualitative  upper limit to the influence of the matter in the equatorial outflow
when maximally enhanced by the long-term change of the stellar mass-loss rate.

The ratio between what appears to be the maximal and the stable radii of influence of the disc matter 
is consistent with a possible increase of the EW of the H$\alpha$ line.
Outer radii of discs in binaries are expected to be truncated by the gravitational influence of their compact companions; at the periastron distances in systems of high eccentricity, and by resonances between the orbital period and the disc gas rotational periods in the low-eccentricity systems \cite{Okazaki2001}. 
\lsi\ is a system between these two cases. 
The effects of the Be star's rotation, which have only recently started to be taken into account, may  modify this conclusion, predicting disc sizes in excess of 10 $R_s$ \cite{Lee2013}.
Assuming the relation between disc size and the EW of the H$\alpha$ by \cite{Grundstrom2006}, and not taking into account rotation effects, 
typical values of the EW  of \lsi\ would lead to an estimation of the disc radius  of the order of the periastron distance
\cite{Grundstrom2007}. 
Simulations indicate that tidal pulls at periastron can lead to the development of large spiral waves in the disc that can extend far beyond the truncation radius and out to the vicinity of the companion (see e.g., \citealp{Okazaki2001}),
promoting accretion \cite{Grundstrom2007}. The $\gamma$-ray data apparently 
provide a window to infer the extent of these waves.

Depending on the period  and dipolar magnetic field, a highly-magnetized neutron star can transition 
between states along the orbital evolution of \lsi, changing its behavior from propeller (near periastron) to ejector (near apastron) along each orbit \cite{Zamanov2001,Torres2012,Papitto2012}. 
These changes of state can be affected by the superorbital variability, since for a larger disc-influence radius, 
the system will remain in the same environment for a longer time \cite{Papitto2012}. The orbital variability is consequently reduced,  
leading to the disappearance of the orbital peak in the power spectrum \cite{Torres2012}.
The data
presented in this report will put the details of this model to the test while
opening the $\gamma$-ray window for studying the discs of Be binaries.

\acknowledgements

The {\emph Fermi} LAT Collaboration acknowledges support from a
  number of agencies and institutes for both development and the
  operation of the LAT as well as scientific data analysis. These
  include NASA and the U.S. Department of Energy (United States); CEA/Irfu and IN2P3/CNRS
  (France); ASI and INFN (Italy); MEXT, KEK, and JAXA (Japan); and
  the K.~A.~Wallenberg Foundation, the Swedish Research Council and
  the National Space Board (Sweden). Additional support from INAF in
  Italy and CNES in France for science analysis during the operations
  phase is also gratefully acknowledged. 
 Additional support of this work comes from 
  grants AYA2012-39303, SGR2009-811, and iLINK2011-0303. 
  DFT was additionally supported by a Friedrich Wilhelm Bessel Award of the Alexander von Humboldt Foundation.

\clearpage

\end{document}